\documentclass{ws-p8-50x6-00}

\def\gtwid{\mathrel{\raise.3ex\hbox{$>$\kern-.75em\lower1ex\hbox{$\sim$}}}}
\def\ltwid{\mathrel{\raise.3ex\hbox{$<$\kern-.75em\lower1ex\hbox{$\sim$}}}}
\def\PRC{{\em Phys. Rev.} C}
\def\APJ{{\em Astrophys. J.}}
\def\SCI{{\em Science}}
\def\NPPS{{\em Nucl. Phys. (Proc. Suppl.)}}
\def\NAT{{\em Nature}}
\def\ASJ{{\em Astron. J.}}
\def\IB{{\em ibid.}}

\begin{document}

\title{The Hot Dark Matter}

\author{David O. Caldwell}

\address{Institute for Nuclear and Particle Astrophysics and Cosmology\\
and\\ Physics Department, University of California, Santa Barbara, CA\\
93106, USA\\E-mail: caldwell@slac.stanford.edu}


\maketitle

\abstracts{
There is a puzzling contradiction: direct observations favor a low-mass-density
universe ($0.2\le\Omega_m\le0.6$), but the only model which fits universe
structure over more than three orders of magnitude in distance scale has a mix
of hot (neutrino) and cold dark matter providing a critical density universe.
Models of an open universe (low $\Omega_m$) or one adding a cosmological
constant ($\Lambda$) to provide a critical energy density ($\Omega_m+
\Omega_\Lambda=1$) have probabilities of $<10^{-3}$.  Two-neutrino dark matter
works better than having the needed $\sim5$ eV of neutrino mass in one species
of neutrino, and this is consistent with the only model which fits all present
indications for neutrino mass: $\nu_\mu\to\nu_\tau$ accounting for the
atmospheric anomaly (with $\nu_\mu$ and $\nu_\tau$ being the hot dark matter),
$\bar\nu_\mu\to\bar\nu_e$ being observed by LSND, and $\nu_e\to\nu_s$
explaining the solar $\nu_e$ deficit.  The LSND/KARMEN results are consistent
with the needed mass of hot dark matter.  Further support for this mass pattern
is provided by the need for the sterile neutrino, $\nu_s$, to make possible
heavy-element nucleosynthesis in supernovae.  It is a fascinating question as
to whether the hot dark matter paradox will be resolved by better measurements
or by the introduction of new physics.}

\section{One-, Two, or Three-Neutrino Dark Matter?}
Since there are about 100/cm$^3$ of neutrinos of each type left over from
the Big Bang, if they have mass they surely are part of the dark matter of
the universe.  While they cannot be the major component of this missing mass,
the neutrinos can have profound effects on universe structure if they have
sufficient mass.  So far there is only evidence for differences in mass
between neutrino types, and with one exception, those differences are so
small that if they are representative of mass values, the neutrinos would
have little effect.  We shall see, however, that there are several types of
evidence that neutrino hot dark matter is quite significant.

The common view, especially among astronomers, has been that if there is
hot dark matter it is due mainly to one neutrino, presumably the $\nu_\tau$.
This would be ruled out if, as fits the Super-Kamiokande data\cite{ref:1}
best, the atmospheric anomalous $\nu_\mu/\nu_e$ ratio is due to
$\nu_\mu\to\nu_\tau$, since the mass-squared difference required is
$\Delta m^2_{\mu\tau}\sim10^{-3}$eV$^2$, whereas the needed neutrino mass is
94 $\Omega_\nu h^2\sim$ eV (with $h$ the Hubble constant in units of 100
km$\cdot\rm s^{-1}\cdot Mpc^{-1}$).  Other processes to explain the
atmospheric results are very unlikely: $\nu_\mu\to\nu_e$ does not fit the
Super-Kamiokande angular distributions, and the CHOOZ $\nu_e$ disappearance
experiment\cite{ref:2} rules out almost all the parameter space;
$\nu_\mu\to\nu_s$ (a sterile neutrino) likely has a problem with the
nucleosynthesis limit,\cite{ref:3} since near maximal mixing is required,
and the Super-Kamiokande results rule it out at the 95\% C.L.

If hot dark matter being one active neutrino is essentially ruled out, what
about using all three active neutrinos?  In this case $\nu_\mu\to\nu_\tau$
explains the atmospheric anomaly, $\nu_e\to\nu_\mu$ (with $\Delta m^2_{e\mu}
\ltwid10^{-5}$ eV$^2$) provides the solar $\nu_e$ deficit, and the three
nearly mass degenerate neutrinos could provide the dark matter.  When this
scheme was first suggested,\cite{ref:4} there was a possible problem with
neutrinoless double beta decay.  While limits on that process have improved,
theoretical ways have been found to ameliorate the problem.  If results from
the LSND experiment\cite{ref:5} are correct, however, three-neutrino dark
matter is also ruled out, since this requires $\Delta m^2_{e\mu}>0.3$ eV$^2$,
making three quite distinct mass differences, necessitating more than three
neutrinos.

That leaves two-neutrino dark matter.  This scheme\cite{ref:4,ref:6} requires
four neutrinos, with the solar deficit explained by $\nu_e\to\nu_s$, both
neutrinos being quite light, the atmospheric effect due to
$\nu_\mu\to\nu_\tau$, which share the dark matter role, and the LSND
$\nu_\mu\to\nu_e$ demonstrating the mass difference between these two
nearly mass-degenerate doublets.  Note that the solar $\nu_e\to\nu_s$ is for
the small mixing angle (or ``just-so'' vacuum oscillation) solution, so
$\nu_s$ does not affect nucleosynthesis.  The original
motivation for this mass pattern preceded LSND and was simply to provide some
hot dark matter, given the solar and atmospheric phenomena.  If LSND is
correct, it becomes the unique pattern.

This neutrino scheme was the basis for simulations\cite{ref:7} which showed
that two-neutrino dark matter fits observations better than the one-neutrino
variety.  The latter produces several problems at a distance scale of the
order of $10h^{-1}$ Mpc, particularly overproducing clusters of galaxies.
Whether the $\sim5$ eV of neutrino mass is in the form of one neutrino
species or two makes no difference at very large or very small scales, but
at $\sim10h^{-1}$ Mpc the larger free streaming length of $\sim5/2$ eV
neutrinos washes out density fluctuations and hence lowers the abundance of
galactic clusters.  In every aspect of simulations done subsequently, the
two-neutrino dark matter has given the best results.  For example, a single
neutrino species, as well as low universe density models, overproduce void
regions between galaxies, whereas the two-neutrino model agrees well with
observations.\cite{ref:8}

\section{Evidence from Universe Structure}
Cosmic microwave background radiation observations of the first Doppler peak
favor the total energy density of the universe having the critical value
($\Omega=1$) of a flat universe.  Such a flat universe has the only time-stable
value of density and is expected in all but rather contrived models of an
early era of exponential expansion, or ``inflation''.  Until recently it has
usually been assumed that $\Omega=\Omega_m=1$; i.e., the energy density is
the matter density, and the universe will expand forever at an ever decreasing
rate.  Now evidence points to $0.3\le\Omega_m\le0.6$, however, based on a
number of observations: high-redshift Type Ia supernovae, evolution of
galactic clusters, high baryon content of clusters, lensing arcs in clusters,
and dynamical estimates from infrared galaxy surveys.  On this basis it has
become popular to assume $\Omega_m\approx0.3$, but $\Omega=1$ through the
addition of a vacuum energy density, often designated as a cosmological
constant, $\Lambda$.  The model with $\Omega_m=0.3$, $\Omega_\Lambda=0.7$
is in trouble with some determinations of the age of the universe, and lensing
measurements require $\Omega_\Lambda<0.74$ at the two standard deviation
level.

The evidence for low $\Omega_m$ does not include a global look at universe
structure.  Gawiser and Silk\cite{ref:9}, however, used all the published
data from the cosmic microwave background and galaxy surveys which covered
three orders of magnitude in distance scale and came to a different conclusion.
They compared the data with ten models of universe structure, but of concern
here are only three of these, the other seven giving extremely poor fits.
In two low-density models the parameters were varied to get best fits,
resulting in $\Omega_m=0.5$, of which the baryons contribute $\Omega_b=0.05$,
and the rest is cold dark matter.  One of these is an open universe model
(OCDM) having $\Omega=\Omega_m=0.5$, and the other ($\Lambda$CDM) has
$\Omega=1$ with $\Omega_\Lambda=0.5$.  The third model (CHDM) has $\Omega_m=1$,
of which $\Omega_\nu=0.2$ is in neutrinos, and $\Omega_b=0.1$ in baryons,
with the main component being cold dark matter, that which was nonrelativistic
at the time it dropped out of equilibrium in the early universe.
The probabilities of the fits were $\rm CHDM=0.09$,
$\rm OCDM=2.9\times10^{-5}$, and $\rm\Lambda CDM=1.1\times10^{-5}$.  If one
dubious set of data is removed, the APM cluster survey (which disagrees with
galaxy power spectra), these probabilities become $\rm CHDM=0.34$,
$\rm OCDM=6.7\times10^{-4}$, and $\rm\Lambda CDM=4.3\times10^{-4}$.

Had it been possible to extend the fit to even smaller scales, the discrepancy
between CHDM and the others would have been even greater, but this is the
non-linear regime requiring simulations.  The CHDM model with two neutrinos
gives an excellent fit\cite{ref:10} to the data at this extended scale,
whereas the others deviate even more strongly than in the linear region.

Clearly there is a serious conflict between the observations indicating a low
value of $\Omega_m$ and the degree of structure in the universe as a function
of distance scale.  Two recent developments raise some doubts about the
present popular interpretation of the data.  The most compelling evidence
for $\Lambda$ comes from the observation of distant
supernovae Ia,\cite{ref:11a} but
recent measurements on nearby SNIa shows\cite{ref:11} they take over two days
longer to reach peak brightness than do distant SNIa.  This indicates that
these may not be the ``standard candles'' required for the conclusions reached,
but rather that evolutionary effects mimic the need for $\Lambda$.  The second
straw in the wind is a geometric measurement of the distance to galaxy
NGC4258 which disagrees with the standard Cepheid ladder of distances by
$\sim15$\%.\cite{ref:12}  This lowering of the distance scale would reduce
the age of the universe, more in line with larger
values of $\Omega_m$.  It is, however, a single determination.  Better
measurements will be available soon, and these may resolve the conundrum,
but if the difference sharpens, this could lead to important new physics.

\section{Evidence from the LSND Experiment}
While new galaxy surveys and cosmic microwave background experiments will be
so precise that they can provide a good measure of neutrino mass and the
number of neutrino types contributing to dark matter, the ultimate answer must
come from laboratory experiments.  Neutrino oscillation experiments can provide
only mass differences, but if the difference is sufficient, the case for
cosmologically significant neutrino dark matter will be settled.  Only one
experiment\cite{ref:5} has given evidence for such dark matter, but so far
the range
of possible mass difference between $\nu_e$ and $\nu_\mu$ is quite large:
0.3 eV$^2\le\Delta m^2_{e\mu}\le10$eV$^2$.

In its 1996 publication,\cite{ref:5} LSND claimed a signal in
$\bar\nu_\mu\to\bar\nu_e$ on the basis that 22 events of the type
$\bar\nu_ep\to e^+n$ were seen, using a stringent criterion to reduce
accidental coincidences between $e^-$ or $e^+$ and $\gamma$ rays mimicking
the 2.2-MeV $\gamma$ from $np\to d\gamma$, whereas only $4.6\pm0.06$ events
were expected.  The probability of this being a fluctuation is
$4\times10^{-8}$.  Note especially that these data were restricted to the
energy range 36 to 60 MeV to stay below the $\bar\nu_\mu$ endpoint and to
stay above the region where backgrounds are high due to the
$\nu_e\/^{12}{\rm C}\to e^-X$ reaction.  In plotting $\Delta m^2$
vs.\ $\sin^22\theta$, however, events down to 20 MeV were used to increase
the range of $E/L$, the ratio of the neutrino's energy
to its distance from the target to detection.  This was done because the plot
employed was intended to show the favored regions of $\Delta m^2$, and all
information about each event was used.  The likelihood analysis applied did
not have a Gaussian likelihood distribution, since its integral is infinite,
but the likelihood contour labeled ``90\%'' was obtained by going
down a factor of 10 from the maximum, as in the Gaussian case.  The
contours in the LSND plot have been widely misinterpreted as confidence
levels---which they certainly are not---because they were plotted along with
confidence-level limits from other experiments.

Recently the difficult, computer-intensive analysis in terms of real
confidence levels has been done.\cite{ref:13}  The likelihood for a grid in
($\sin^22\theta$, $\Delta m^2$) space, including backgrounds, has been computed
and compared with numerous Monte Carlo experiments to obtain a 90\% confidence
region.  While the equivalency varies from point to
point in the $\Delta m^2-\sin^22\theta$ plane, a typical value for the 90\%
confidence level is down a factor of 20 from the likelihood
maximum.  Thus the LSND allowed regions are considerably broader in
$\sin^22\theta$ than in the plots published so far, and other experiments
are less constraining of allowed $\Delta m^2$ regions.

The confusion of comparing likelihood levels for LSND with confidence levels
from other experiments may be exacerbated by using the 20--36 MeV region
for the LSND data.  While this higher background energy range makes some
difference for the 1993--5 data, it could have had an appreciable effect for
the parasitic 1996--7 runs, which were at a low event rate.  This decreased
the ratio of signal/background events, since the main background is from
cosmic rays.  This could raise the low end of the supposed signal energy
spectrum, making the higher $\Delta m^2$ values desirable for dark matter
appear less likely.

\begin{figure}[t]
\epsfxsize=12cm
\epsfbox{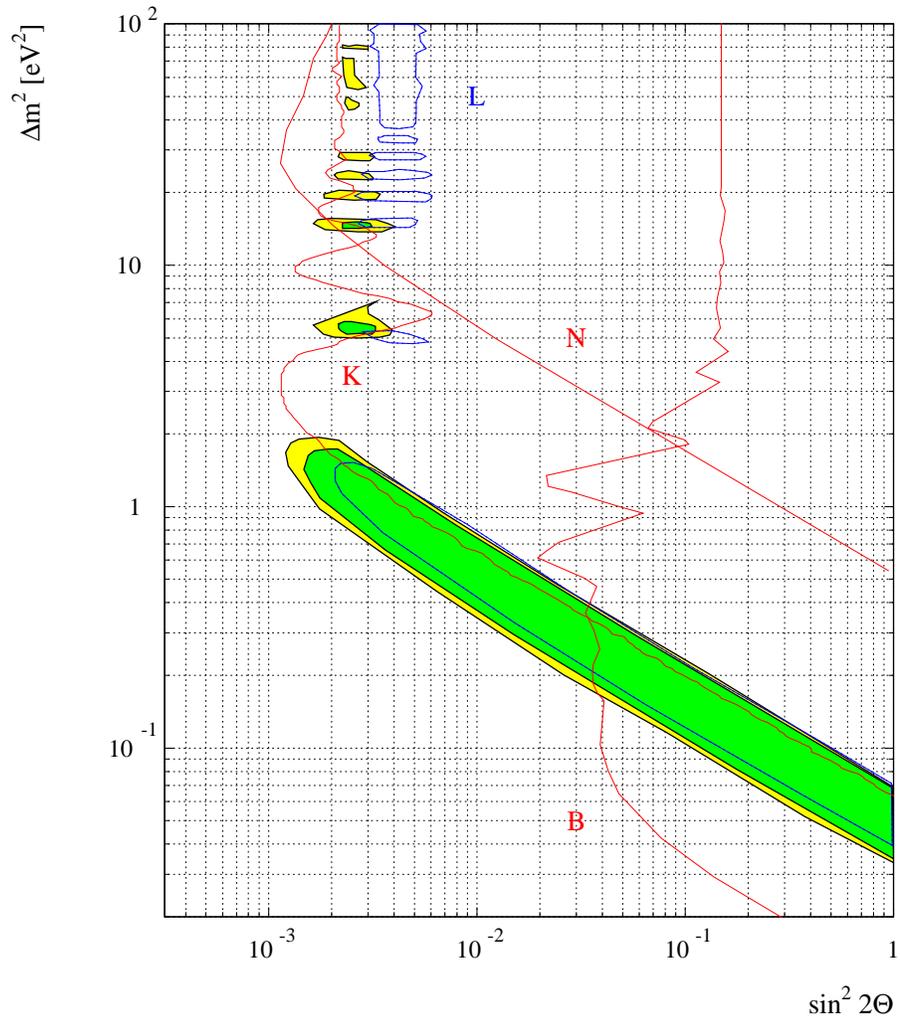} 
\caption{Filled in areas are 90\% and 95\% confidence regions based on
the product of the KARMEN and LSND Feldman-Cousins likelihood ratios.  Also
shown is the Feldman-Cousins 90\% confidence region for LSND alone (``L'').
Left of the ``K'', ``N'', and ``B'' curves are exclusion regions of
KARMEN, NOMAD, and Bugey.\label{fig:1}}
\end{figure}

Nevertheless, when a joint analysis is made of the LSND and KARMEN\cite{ref:13}
experiments even using the 20--36 MeV range for LSND, the region around 6
eV$^2$ is as probable as the banana-shaped region at lower $\Delta m^2$, as
shown in Fig.~\ref{fig:1}.  Frequently theorists consider only the latter,
whereas the $\nu_\mu\to\nu_e$ LSND data favors the higher mass region.  Of
course the $\nu_\mu\to\nu_e$ data,\cite{ref:5} which uses $\nu_\mu$ from
$\pi^+$ decay-in-flight and detects $\nu_e$ by $\nu_e\/^{12}{\rm C}\to
e^-X$ has higher backgrounds and hence much poorer statistics than the
$\bar\nu_\mu\to\bar\nu_e$ with $\bar\nu_\mu$ from $\mu^+$ at rest.  In
addition to the $\Delta m^2$ issue, the important point of Fig.~\ref{fig:1}
is that although the KARMEN data are consistent with background, the joint
analysis of the $\bar\nu_\mu\to\bar\nu_e$ data from the two experiments shows
an appreciable region for a signal.  KARMEN is continuing to take data, and
LSND should have an improved analysis available soon.

\section{Evidence from Supernova Nucleosynthesis}
Support for the double doublet of neutrinos with sufficient mass in two of the
neutrinos to provide significant hot dark matter comes from an unusual source:
the creation of heavy nuclei by supernovae.  Initially the reverse appeared
to be the case, since this $r$ process of rapid neutron capture, which occurs
in the outer neutrino-heated ejecta of Type II supernovae, seemed to place
a limit on the mixing of $\nu_\mu$ and $\nu_e$.  Energetic $\nu_\mu$
($\langle E\rangle\approx25$ MeV) coming from deep in the supernova core
could convert via an MSW transition to $\nu_e$ inside the region of the
$r$-process, producing $\nu_e$ of much
higher energy than the thermal $\nu_e\ (\langle E\rangle\approx11$ MeV).  The
latter, because of their charge-current interactions, emerge from farther out
in the supernova where it is cooler.  Since the cross section for $\nu_en\to
e^-p$ rises as the square of the energy, these converted energetic $\nu_e$
would deplete neutrons, stopping the $r$-process. 
Calculations\cite{ref:14} of this effect limit $\sin^22\theta$ for
$\nu_\mu\to\nu_e$ to $\ltwid10^{-4}$ for $\Delta m^2_{e\mu}\gtwid2$ eV$^2$, in
conflict with compatibility between the LSND result and a neutrino component of
dark matter.

Since the work of reference 15, serious problems have been found with the
$r$ process itself.  First, recent simulations have revealed the $r$-process
region to be insufficiently neutron-rich, since about $10^2$ neutrons is
required for each seed nucleus, such as iron.  This was bad enough, but the
recent realization of the full effect of $\alpha$-particle formation has
created a disaster for the $r$ process.\cite{ref:15}  At a radial region
inside where the $r$ process should occur, all available protons swallow
up neutrons to form the very stable $\alpha$ particles, following which
$\nu_en\to e^-p$ reactions reduce the neutrons further and create more protons
which make more $\alpha$ particles, and so on.  The depletion of neutrons by
making $\alpha$ particles and by $\nu_en\to e^-p$ rapidly shuts off the
$r$ process, and essentially no nuclei above $A=95$ are produced.

What is required to solve this problem is to remove the $\nu_e$ flux at the
$r$ process site, but there still has to be a very large $\nu_e$ flux at a
smaller radius for material heating and ejection.  This apparent miracle
can be accomplished\cite{ref:16} if there is (1) a sterile neutrino, (2)
approximately maximal $\nu_\mu\to\nu_\tau$ mixing, (3) small $\nu_\mu\to\nu_e$
mixing, and (4) an appreciable ($\gtwid2$ eV$^2$) mass-squared difference
between $\nu_s$ and the $\nu_\mu$--$\nu_\tau$.  This is precisely the
neutrino mass pattern required to explain the solar and atmospheric anomalies
and the LSND result, plus providing some hot dark matter!

Such a mass-mixing pattern creates two level crossings.  The inner one,
which is outside the neutrinosphere (beyond which neutrinos can readily
escape) is near where the $\nu_{\mu,\tau}$ potential $\propto(n_{\nu_e}-n_n/2)$
goes to zero.  Here $n_{\nu_e}$ and $n_n$ are the numbers of $\nu_e$ and
neutrons, respectively.  The $\nu_{\mu,\tau}\to\nu_s$ transition which occurs
depletes the dangerous high-energy $\nu_{\mu,\tau}$ population.  Outside
of this level crossing, another occurs where the density is appropriate for a
matter-enhanced MSW transition corresponding to whatever $\Delta m^2_{e\mu}$
LSND is observing.  Because of the $\nu_{\mu,\tau}$ reduction at the first
level crossing, the dominant process in the MSW region reverses from the
deleterious $\nu_{\mu,\tau}\to\nu_e$, becoming $\nu_e\to\nu_{\mu,\tau}$ and
dropping the $\nu_e$ flux going into the $r$-process region.  For an
appropriate value of $\Delta m^2_{e\mu}$, the two level crossings are
separate but sufficiently close so that the transitions are coherent.  Then
in the limits of adiabatic transitions and near maximal $\nu_\mu$--$\nu_\tau$
mixing, the neutrino flux emerging from the second level crossing is 1/4
$\nu_\mu$, 1/4 $\nu_\tau$, and 1/2 $\nu_s$, with no $\nu_e$ at all.
Calculations show the transitions to be adiabatic, and the atmospheric
observations require near maximal mixing, so the $\nu_e$ flux is certainly
sufficiently depleted to allow a successful $r$ process, especially as the
$\bar\nu_e$ flux is unaffected, so that $\bar\nu_ep\to e^+n$ enhances the
neutron number.  It should be emphasized that this mechanism is quite robust,
not depending on details of the supernova dynamics, especially as it occurs
quite late in the explosive expansion.

It is essential that the two level crossings be in the correct order, and this
provides a requirement on $\Delta m^2_{e\mu}$, since the MSW transition
depends on density and hence on radial distance from the protoneutron star.
Detailed calculations have been made for $\Delta m^2_{e\mu}\sim6$ eV$^2$,
which works very well.  Possibly $\Delta m^2_{e\mu}$ as low as 2 eV$^2$ or
maybe even 1 eV$^2$ would work, but that is speculative.  At any rate, the
mass difference needed in this scheme, which is the only one surely consistent
with all manifestations of neutrino mass and which rescues the $r$
process,\cite{ref:17} implies appreciable hot dark matter.

\section{Conclusions}
A neutrino component of dark matter appears very probable, both from the
astrophysics and particle physics standpoints.  Despite abundant evidence for
$\Omega_m<1$, the one model which fits universe structure has $\Omega_m=1$,
with 20\% neutrinos and most of the rest as cold dark matter.  Open universe
and low-density models with a cosmological constant give extremely bad fits.
This conflict should be the source of future progress, but since there are
$10^2/\rm cm^3$ of neutrinos of each active species left over from the early
universe, the ultimate answer on neutrino dark matter will come from
determinations of neutrino mass.  While the solar and atmospheric evidences
for neutrino mass are important, the crucial issue is the much larger
mass-squared difference observed by the LSND experiment.  In the mass region
needed for dark matter, no other experiment excludes the LSND result, and a
joint analysis of the LSND and KARMEN experiments shows this region has
good probability.

The resulting mass pattern, $\nu_e\to\nu_s$ for solar, $\nu_\mu\to\nu_\tau$
for atmospheric, and $\nu_\mu\to\nu_e$ for LSND, requires a sterile neutrino
and provides two-neutrino ($\nu_\mu$ and $\nu_\tau$) dark matter.  This
form of dark matter fits observational data better than the one-neutrino
variety.  Furthermore, the four-neutrino pattern, and especially the sterile
neutrino, provides a robust way to make possible the production of heavy
elements by supernovae.

\section*{Acknowledgments}
This work was supported in part by the U.S.~Department of Energy.  Portions
of this paper were done in enjoyable past collaborations with R.N.~Mohapatra,
J.R.~Primack, G.M.~Fuller, and Y.-Z.~Qian.  Assistance from S.J.~Yellin is
much appreciated.


\begin{thebibliography}{99}
\bibitem{ref:3}At the time of writing there is considerable controversy
about the nucleosynthesis limit, not only regarding concordance of
light-element abundances, but also about the possible effects of neutrino
particle-antiparticle asymmetries.
\end{thebibliography}
\end{document}